\documentclass{aa}
\usepackage{txfonts}
\usepackage{graphicx}
\usepackage{natbib}
\usepackage{epsfig}

\begin{document}
\title{Mass Loss, Destruction and Detection of Sun-grazing \& -impacting Cometary Nuclei}

\author{J.C. Brown\inst{1,4,5}, H.E. Potts\inst{1}, L.J. Porter\inst{1,2} \& G. le Chat\inst{1,3}}

\institute{$^{1}$ School of Physics \& Astronomy, University of
Glasgow, G12 8QQ, UK.
 \email{john.brown@glasgow.ac.uk}\\
 $^{2}$ Max-Planck-Institut f\"{u}r Astrophysik, D-85748 Garching, Germany\\ $^{3}$ LESIA, Observatoire de Paris, CNRS, UPMC, Universit\'{e} Paris Diderot; 92190 Meudon, France\\$^{4}$ NCAR High Altitude Observatory, Boulder, CO 80303, USA\\$^{5}$Astrophysics Research Group. Trinity College,  Dublin (TCD), Ireland\\}
\offprints{J.C. Brown, \email{john@astro.gla.ac.uk}}

\authorrunning{Brown {\it et al.}}
\titlerunning {Sun grazing and Impacting Comets}
 \date{\today}

\abstract {Sun-grazing comets almost never re-emerge, but their sublimative destruction near the sun has only recently been observed directly, while chromospheric impacts have not yet been seen, nor impact theory developed.}
{We seek simple analytic models of comet destruction processes near the sun, to enable estimation of observable signature dependence on original incident mass $M_o$ and perihelion distance $q$.}
{Simple analytic solutions are found for $M(r)$ versus $q$ and distance $r$ for insolation sublimation and, for the first time, for impact ablation and explosion.}
{Sun-grazers are found to fall into three ($M_o,q$) regimes: sublimation-, ablation-, and explosion-dominated. Most sun-grazers have $M_o$ too small ($< 10^{11}$ g) or $q$ too large ($>1.01R_\odot$) to reach atmospheric densities ($n>2.5\times 10^{11}$/cm$^3$) where ablation exceeds sublimation. Our analytic results for sublimation are similar to numerical models.
For $q<1.01R_\odot, M_o>10^{11}$ g, ablation initially dominates but results are sensitive to nucleus strength $P_c=10^6P_6$ dyne/cm$^2$ and entry angle $\phi$ to the vertical.  Nuclei with $M_o \preceq 10^{10}(P_6\sec\phi)^3$ g are fully ablated before exploding, though the hot wake itself explodes. For most sun-impactors $\sec\phi \gg 1$ (since $q\sim r_*$), so for $q$ very close to $r_*$ the ablation regime applies to moderate $M_o \sim 10^{13-16}P_6^3$ g impactors unless $P_6\preceq
0.1$. For higher masses, or smaller $q$, nuclei reach densities $n> 2.5\times 10^{14}P_6$/cm$^3$ where ram pressure causes catastrophic explosion.}
{Analytic descriptions define ($Mo,q$) regimes where sublimation, ablation and explosion dominate sun-grazer/-impactor destruction. For $q\prec 1.01R_\odot, M_o\succeq 10^{11}$ g nuclei are destroyed by ablation or explosion (depending on $M_o\cos^3\phi/P_c$) in the chromosphere, producing flare-like events with cometary abundance spectra. For all plausible $M_o,q$ and physical parameters, nuclei are destroyed above the photosphere.}

\keywords{comet; sun; impact; sun-grazing; radiation; collisions}
\maketitle

\footnote[1]{This paper is dedicated to the memories of: Brian G.Marsden, world expert on minor bodies of the solar system and an irreplaceable friend and colleague; Gerald S. Hawkins who introduced me to the joys of this field in my (JCB's) first real (radar meteor) research experience at HSRAO/CfA in the summer of 1967.}

\section{Introduction}
\label{Introduction}

Close sun-grazing comets are discovered almost daily by white light coronographs (e.g. SoHO LASCO). Most are small and fully sublimated by insolation at a few solar radii $R_\odot$
 while almost none have re-emerged (Marsden 2005). The majority have perihelion distances $q$ well above $R_\odot$ but some have $q\approx$ or $\prec R_\odot$ (e.g. Marsden 2005 and http://www.minorplanetcenter.org/mpec/RecentMPECs.html.) The death of a comet at $r\sim R_\odot$ has been seen directly only very recently (Schrijver et al 2011) using the SDO AIA XUV instrument. This recorded sublimative destruction of Comet C/2011 N3 as it crossed the solar disk very near perihelion $q=1.139R_\odot$. The next challenge in studying the demise of close sun-grazers will be to catch one of the even rarer cases of chromospheric impact ($q<1.01R_\odot$, $M_o>10^{11}$ g). As anticipated by Weissman (1983) and shown below, these undergo explosive destruction in the dense chromosphere. Understanding the destruction processes, and their radiation signatures, are essential steps in searching for and modeling these.

The processes leading to sublimation of the icy conglomerate mix (Whipple 1950) of cometary nuclei, and in some cases their splitting and fragmentation, were considered by  Huebner (1967), Weissman and Kieffer (1981),
Weissman (1983), Iseli et al. (2002), Sekanina (2003) and others. These models essentially solve for the insolative sublimation mass loss rates of icy-conglomerate mixes (the dust being carried away in the flow of these evolved components). They
variously allow for the complicating factors of rotation, albedo, insulating surface dust layers, radiative cooling, interior thermal conduction, and fragmentation by tidal, thermal and volatile explosion effects. Huebner (1967) and Iseli et al. (2002), for example, found that, for high sublimation rates near the sun, these effects were secondary, the mass loss being reasonably approximated by a pure sublimation description : mass loss rate = heating rate/latent heat. Using SoHO data, Sekanina (2003) addressed in detail how mass loss rates and fragmentation relate to cometary light curves via atomic line emission (e.g. by sodium) and by dust scattering of sunlight, though emphasising that most of the mass remains in a primary fragment. None of these studies considered ablation or ram-pressure driven explosion due to solar atmospheric impact (which we show below are negligible till $r\preceq1.01R_\odot$) though  Weissman (1983) had remarked {\it "ultimate destruction of the nucleus [of sun-impacting comets] likely results from the shock of hitting the denser regions of the solar atmosphere"}. It is well known in the planetary physics community that ablation and explosion are central processes in comet-planetary atmosphere impacts (e.g. Carslon et al. 1997)

 Here we revisit the theory of sun-grazer sublimation then develop the first estimates of the much higher rates of mass loss by ablation for the rarer cases which reach  $\preceq1.01R_\odot$. This regime (Sections 3.4, 6.1, 6.2) resembles that of comet-planetary impacts though with some differences. One is that almost all sun-grazing comets belong to secondary comet groups (mostly Kreutz) formed from primordial comets. A significant number of the latter must have $q<R_\odot$ (Hughes 2001) and large enough mass ($M_o\succeq 10^{11}$ g - see Section 5, Eqn. (\ref{Mominq})) to survive sublimation down to the intense ablation/explosion regime. However only a small fraction of group comets (Biesecker et al. 2003, Knight et al 2010) come that close or are that massive, larger sun-grazers mostly having $q\succeq 1.5 R_\odot$. Comet C/2010 E6 (STEREO) discussed by Raftery et al. (2010) came close,  having $q\approx 1.02 R_\odot$ while the destruction of Comet C/2011 N3 seen by SDO (Schrijver et al 2011) on the solar disk was solely by sublimation as it had $q\approx 1.14R_\odot$. Most of the few group comets reaching the ablation/explosion regimes will have small mass and very shallow incidence angles (since $q\approx R_\odot$) with implications for the relative importance of ablation and explosion. By analogy with models of planetary atmospheric impacts (e.g. Carlson et al. 1997) , chromospheric impactors are expected to lose mass, momentum and kinetic energy very rapidly, the soaring temperature and pressure leading to destructive detonation of the nucleus and its wake. This 'air burst' should be followed by explosive expansion of the resulting 'fireball' of cometary and solar atmospheric material. The total masses and energies of these explosive chromospheric events are in roughly the same range as those of solar magnetic flares so they could be termed 'cometary flares'.

  Section 2 discusses relevant nucleus parameters values while Section 3 estimates the relative importance and time/distance-scales of nucleus mass loss, explosion, and deceleration processes. We then consider in some detail approximate analytic treatment of mass loss (Section 4) purely by sublimation (Section 5) and by ablation and explosion (Section 6).

\section{Comet Nucleus Parameters}
\subsection{Masses and Densities}

In the past there was wide range of opinions over the masses $M_o$ and mean mass
densities $\rho$ of the icy-conglomerate mix thought to comprise comet nuclei, and
on how these quantities varied between different comets (Whipple 1950). We simply take the incident mass $M_o$ to be a parameter ranging from  the faintest comets detected almost daily to the largest super-massive comets. Concerning the density $\rho$, MacLow and Zahnle (1994) said {\it the density of comets is the stuff of speculation [probably] 0.3 - 1.0 g/cm$^3$ but with extreme claims of 0.01-5}. In his sublimation modeling Sekanina (2003) used $\rho=0.15$ g/cm$^3$. More recently, however, Weissman and Lowry (2008) critically discussed results from a range of methods and data, including Deep Impact. All results have a considerable error range, but they conclude that $\rho =0.6\pm 0.2$ g/cm$^3$ is the most likely range for the comets they studied (cf Richardson et al. 2007). Here we therefore also take $\rho$ as a parameter, and express it relative to that of (water) ice $\rho_{ice}=0.9$ g/cm$^3$ via the dimensionless scaling parameter

\begin{equation}
\label{deftilderho}
\tilde{\rho}=\rho/\rho_{ice}
\end{equation}
In discussing results we mainly use a mean value $\tilde{\rho}=0.5$ ($\rho \approx 0.45$ g/cm$^3$)
though showing how results, such as for $M_o$, can be scaled for other values (see Eqn. (\ref{effectivemass}) and Figure 1).

\subsection{Latent heat of sublimation}
The major nucleus parameter determining its rate of mass loss for power input ${\cal P}$ (erg/s) is the mean latent heat ${\cal L}$ (erg/g) of sublimation/ablation for the cometary ice-conglomerate mix. The value for water ice is about $2.6\times10^{10}$ erg/g. This is of order the relevant value when the mass lost is in the form of water vapor whose pressure sweeps dust and other non-volatiles away with it. However, in situations of intense ablation and explosion, all components are eventually vaporized and the relevant value is the weighted mean over all mass components, including volatile and refractory ones. Sekanina (2003) finds values for what he terms 'effective latent heat of erosion'  to be around 0.3 of the water ice value. However Huebner et al's. (2006) mass composition [Silicates : organics : carbonaceous: ices ] = [0.26:0.23:0.086: 0.426] suggests refractory silicates to be the most important non-ice component. Mendis (1975) give ${\cal L}=2.30\times10^{10}$ erg/g for silicates, which is close to that of water ice, while organics will lower the mean slightly. Chyba et al (1993) adopted $2.3\times 10^{10}$ erg/g for comets, $5\times 10^{10}$ erg/g for carbonaceous and $8\times 10^{10}$ erg/g for stone/iron bodies and just $10^{11}$ erg/g even for solid iron. Here we use the water ice value for ${\cal L}$ as a reasonable first approximation, but include a dimensionless scaling parameter
\begin{equation}
\label{deftildecalL}
\tilde{{\cal L}}={\cal L}/{\cal L}_{ice}
\end{equation}
to allow application to other ${\cal L}$ values though adopting $\tilde{{\cal L}}\approx 1, ({\cal L}\approx {\cal L}_{ice})$, in most numerical evaluations. To convert to other values see Figure 1 and Eqn. (\ref{effectivemass}).
\subsection{Physical Strength and Sound Speed}
The effects on the nucleus of the pressure associated with the processes driving sublimation/ablation depend on how strong the nucleus material is and how fast pressure waves propagate through its volume. In the initial stages the relevant strength is that of the loose low density ice-conglomerate mix, estimated by Zahnle and MacLow (1994) to be a pressure ($\approx$ energy density) of $P_c\approx 10^6$ dyne/cm$^2(\equiv$ erg/cm$^3)$ while values discussed by Chyba et al (1993) range from $10^8$ dyne/cm$^2$ for very solid asteroid-like bodies to $10^5$
dyne/cm$^2$ for fluffy snowy structures. The last value is also that estimated by Richardson et al (2007) for Comet 9P/Tempel 1 using Deep Impact data, while tidal disruption data suggest the possibility of even lower values for some comets. Here  we allow consideration of a range of values by parametrically setting $P=10^6P_6$ dyne/cm$^2$ and specifically discuss the range  $P_6=0.1-1$. For the sound speed, Zahnle and MacLow (1994) suggest $c_s\approx 1-3\times 10^5$ cm/s for the initial state nucleus material. Values for a range of more or less solid materials do not differ greatly.
\subsection{Nucleus shape and 'size'}
  For a given density the mass loss, deceleration and explosion rates for a nucleus depend on its volume $V$ and effective area $A$ which depend on its size scale $'a'$ and shape. Since nuclei have irregular and uncertain shapes, as evidenced by images and light curves, rather than treat precisely the hypothetical case of a sphere, we will use $A\approx a^2, V \approx a^3$ for an effective mean size $a$. This formulation gives $A$ and $V$ correctly to within 10\% for the case of a sphere of radius $b$ if we adopt $a\approx 1.67b$. We use this later when referring to spherical cases for illustration (Sections 5.2.2 and 5.3).

\section{Rough Comparison of Nucleus Mass Loss, Deceleration and Explosion Processes}

\subsection{'Evaporative' Mass Loss versus Deceleration}
Whether mass loss from the nucleus is by sublimation or by ablation, the total energy needed to vaporize the whole nucleus is ${\cal E}_{vap}=M_o{\cal L}$ while the energy needed to stop it is ${\cal E}_{kin}\approx M_ov_\odot^2/2$. The ratio is tiny, ${\cal E}_{vap}/{\cal E}_{kin} \approx 2{\cal L}/v_\odot^2= 4\times 10^{-5}\tilde{\cal L}$ for the solar escape speed

\begin{equation}
\label{defvodot}
v=v_\odot=(2GM_\odot/R_\odot)^{1/2}\approx 618~{\rm km/s}
\end{equation}
with $G,M_\odot,R_\odot$ the gravitational constant, solar mass, and solar radius respectively. Thus total vaporization can occur well before the nucleus decelerates significantly, though ram pressure can result in explosive destruction dominating.
Equivalently, to conserve momentum in slowing down, the nucleus must encounter an atmospheric mass comparable to $M_o$. In doing so it absorbs far more energy than needed to vaporize it. The mass per unit area (g cm$^{-2}$) of a comet nucleus is $\Sigma_c\approx M_o/a^2=M_o^{1/3}\rho^{2/3}=10^4(M_o/10^{12})^{1/3}(\tilde{\rho})^{2/3}$ g cm$^{-2}$ while that of the sun's atmosphere even down to the photosphere is only $\Sigma_\odot \approx 1$ g cm$^{-2}$. Consequently, unless they explode, increasing the deceleration, only objects of $<10^3$ g or so would be much decelerated by the mass of the sun's atmosphere down to the photosphere though, in practice, they would be vaporized earlier by sublimation and/or ablation. Note also that, at infall speed $v_\odot$, in the frame of the nucleus a solar atmospheric proton has kinetic energy $\sim$ 2 keV which is enough to knock off about 4000 water molecules, or about $10^5$ times its own mass.

This situation of initial ablative mass loss without significant deceleration is paralleled by the dynamics in planetary atmospheres of meteors - e.g. McKinley (1961), Kaiser (1962) - and at least of the initial (high altitude) stages of comet-planet impacts - e.g. Shoemaker Levy 9 with Jupiter - cf. Section 6.

In the sublimation regime the radiation has negligible effect on the nucleus speed, even though it delivers a large amount of energy. Specifically the ratio of the radiation pressure force $F_{rad}$ to the gravitational is (ignoring the correction factor 1.0-2.0 for albedo) $ \frac{F_{rad}}{F_{grav}}\preceq \frac{{10^{-4}}}{a(cm)\tilde{\rho}}$. The radiation pressure $F_{rad}/c \simeq 2$ dyne/cm$^2$ is also tiny compared to the nucleus strength $P_c$ so causes no direct explosion effect. It is also straightforward to show that, during perihelion passage, the rocket effect of mass loss leaving the nucleus anisotropically has negligible effect on $v$ during vaporization. This is because the very high nucleus $\rho$ value implies very low mass loss speed $u$ and hence momentum flux. Consequently, the velocity of the nucleus during its vaporization is well approximated by that of a gravitational parabolic orbit,  viz., in orbital plane polar coordinate ($r,\theta$) and for perihelion distance $q$
\begin{eqnarray}
\label{veloccomps}
v_\theta =& r\dot{\theta}=& v_\odot \left(\frac{R_\odot}{r}\right)^{1/2}\left(\frac{q}{r}\right)^{1/2}\nonumber\\
v_r=&\dot{r}=& v_\odot \left(\frac{R_\odot}{r}\right)^{1/2}\left(1-\frac{q}{r}\right)^{1/2}\nonumber\\
v=&(v_\theta^2+v_r^2)^{1/2}=& v_\odot\left(\frac{R_\odot}{r}\right)^{1/2}
\end{eqnarray}
\subsection{Deceleration of sublimated/ablated mass}
While, throughout the total vaporization lifetime of the nucleus,
its velocity is well described by Eqns. (\ref{veloccomps}),
this is not true of the material it loses.
Small particles leaving the nucleus do not obey
Eqns. (\ref{veloccomps}), non-gravitational accelerations on
them being very important. These include $F_{rad}$ on sub-micron particles
and atmospheric drag on atoms and ions. The Coulomb collisional stopping column density for a proton of speed $v_\odot\approx 618$ km sec$^{-1}$ ($\sim 2$ keV) in
a hydrogen plasma is around ${\cal N}_s=10^{15}$ cm$^{-2}$ (Emslie 1978).
Thus the stopping distance at number density $n$ is
$d ({\rm cm}) \approx {\cal N}/n\approx 10^{15}/n$, i.e. just 1 km in the
corona and 0.1 mm in the photosphere. So ablated dust
and ions stop abruptly and form an exploding wake as they blend with and heat the
atmosphere, creating large local enhancements of heavy element abundances.

\subsection{Insolation Sublimation case $q\succeq r_*, M_o\preceq 10^{11}$ g}
Very near the sun, the timescale for sublimation of the whole mass is  $\tau_{sub}\succeq M_o{\cal L}/a_o^2{\cal F}_\odot \approx 4\times 10^3M_{12}^{1/3}\tilde{\rho}^{2/3}\tilde{{\cal L}}$ s, where ${\cal F}_\odot =6\times 10^{10}$ erg/cm$^2$/s is the bolometric photospheric energy flux and $M_{12}=M_o/10^{12}$ g. The corresponding distance scale is roughly $d_{sub}\approx v_\odot\tau_{sub}\approx 3.5R_\odot\times M_{12}^{1/3}\tilde{\rho}^{2/3}\tilde{{\cal L}}$ cm as first seen directly by Schrijver at al. (2011). Total sublimation occurs in a close perihelion passage roughly for $d_{sub}\preceq R_\odot$ which is the case for masses $M_o\preceq 3\times 10^{10}/ \tilde{\rho}^2\tilde{{\cal L}}^3$ g or about $10^{11}$g for
$\tilde{\rho}= 0.5,\tilde{{\cal L}}=1$. This shows the  majority of close sun-grazers ($q\succeq1.01R_\odot$) to have  $M_o \preceq 10^{11}$ g  as they never re-emerge.
\subsection{Impact ablation and ram pressure explosion cases}
\subsubsection{Background}
The problems of comet, asteroid, and even large meteoroid  impact with planetary atmospheres closely parallel those of a solar impact, though the parameter regimes are rather different, and there is no planetary equivalent of insolation. Unlike the solar case, planetary impacts  have been addressed in detail, early work including that of Revelle (1979) and others. Progress was greatly accelerated in anticipation, and in the aftermath, of the collisions of fragments of Comet Shoemaker-Levy 9 with Jupiter in July 1994. Many authors (e.g. Chyba et al 1993, Chevalier and Sarazin 1994, Zahnle and MacLow 1994) developed semi-analytic and numerical models to predict what should be expected of these impacts. Others, notably  MacLow and Zahnle (1994), Field and Ferrara (1995) and Carlson et al. (1995, 1997), developed models further by drawing on actual event data. In particular the Carlson et al. (1995, 1997) 'heuristic model' combined data analysis, numerical simulations, and observational inputs. They identified and addressed both the 'bolide initial phase' (which we have called the detonation or airburst) and the subsequent 'fireball' exploding out and up from the nucleus and its wake. Detailed observations
of the G-fragment impact were synthesised with their numerical simulations of the fireball (e.g. their Figure 7). Chyba et al. (1993), Zahnle and MacLow (1994) and MacLow and Zahnle (1994) showed that the airburst phase is dominated by ablation only in the upper atmosphere, and below that by catastrophic pressure-driven disruption. In the Jupiter case, the latter is true for all impacting masses except those very much smaller than that ($\approx 10^{15}$ g) of the Shoemaker-Levy 9 fragments in 1994. We summarize here the essence of the arguments leading to these conclusions so that we can adapt them to the somewhat different conditions of solar impacts.
\subsubsection{Rough comparison of the importance of impact processes}
The relative importance of ablation, explosion and deceleration in the initial destruction of the nucleus depends on atmospheric scale height $H$ as well as on the nucleus mass, speed and entry angle. They can be expressed in terms of the atmospheric column densities $\cal N$(cm$^{-2})=\int_s nds$ traversed {\it along the path} $s$ of the nucleus after which each processes is complete. (The vertical column density $N={\cal N}\cos \phi$ for constant path angle $\phi$ to the vertical) .
First, to deliver enough energy to ablate the whole mass (without expansion) requires an encounter with column density ${\cal N}_{abl}$ such that ${\cal N}_{abl}\mu m_pv_\odot^2/2\approx {\cal L}M_o/a_o^2$
where $m_p$ is the proton mass and $\mu$ the mean mass present per H (atoms and ions) in units of $m_p$. This implies

\begin{eqnarray}
\label{Nabl}
{\cal N}_{abl} ({\rm cm}^{-2})&=N_{abl}\sec \phi\approx {\cal L}M_o^{1/3}\rho^{2/3}/(\mu m_pv_\odot^2/2)&  \nonumber\\
&\approx 6\times 10^{22}M_{12}^{1/3}\tilde{{\cal L}}\tilde{\rho}^{2/3}&
\end{eqnarray}
Total deceleration of the nucleus (without expansion) requires ${\cal N}_{dec}$ such that ${\cal N}_{dec}\mu m_p \approx M_o/a_o^2$ implying
\begin{equation}
\label{Ndec}
{\cal N}_{dec} ({\rm cm}^{-2})\approx \frac{M_o^{1/3}\rho^{2/3}}{\mu m_p}\approx 4\times 10^{27}M_{12}^{1/3}\tilde{\rho}^{2/3}\approx 3\times 10^{27}M_{12}^{1/3}
\end{equation}
which is $\gg {\cal N}_{abl}$.
For internal strength (energy density) $P_c=10^6P_6$ dyne/cm$^2$, nuclei will undergo hydrodynamic flow once they reach the depth where atmospheric ram pressure $\mu m_pv_\odot^2/2$ approaches the value $P_c$. This implies an atmospheric proton density $n \succeq n_{pres}=n_{**}\approx 2P_c/\mu m_p v_\odot^2$. The corresponding column density ${\cal N}_{pres}$ along the path for atmospheric scale height $H=10^8H_8$ cm is

\begin{eqnarray}
\label{Npres}
{\cal N}_{pres} ({\rm cm}^{-2})&=n_{pres}H\sec\phi\approx \frac{2P_cH \sec \phi}{\mu m_p v_\odot^2}& \nonumber\\ &\approx 2.4\times 10^{22}P_6 H_8\sec \phi&
\end{eqnarray}
The ratio of critical $\cal N$  for ablation end to ram pressure onset is (Eqns. (\ref{Nabl}, \ref{Npres}))

\begin{equation}
\label{NablovrNPres}
\frac{{\cal N}_{pres}}{{\cal N}_{abl}}\approx \frac{0.4P_6 H_8\sec\phi}{M_{12}^{1/3}\tilde{{\cal L}}\tilde{\rho}^{2/3}}\approx \frac{0.7P_6 H_8\sec\phi}{M_{12}^{1/3}}
\end{equation}
where the last expression is for $\tilde{\rho}=0.5,\tilde{{\cal L}}=1$. There is thus a maximum mass $M_{abl}^{max}(\phi)$ above which destruction of the nucleus is driven by explosion rather than ablation, namely (setting ratio (\ref{NablovrNPres}) to unity)

\begin{eqnarray}
\label{MAblmax}
M_{abl}^{max}(\phi_*) & =M_{**}\sec^3\phi\approx 6\times 10^{10}\left[\frac{P_6 H_8\sec\phi}{\tilde{{\cal L}}\tilde{\rho}^{2/3}}\right]^3 {\rm g}& \nonumber\\ &\approx 2.4\times 10^{11}[P_6 H_8\sec\phi]^3{\rm g}&
\end{eqnarray}
the last value again for $\tilde{\rho}=0.5,\tilde{{\cal L}}=1$. (Note - In the rest of this paper we use subscripts $_*$ and $_{**}$ respectively for values of quantities at the points where ablation first exceeds sublimation, and $_{**}$ where explosion first becomes dominant.)

Thus although ram pressure sets in at a lower $n$ for the sun than for Jupiter (because $v$ is higher) the same $v$ dependence occurs in the ablation expression so the ratio (\ref{NablovrNPres}) just scales as $1/H$ and reaches unity only at a higher $n$ in the solar case. For solar
$H_8\sim 0.5$ we find $n_{pres}=n_{abl}^{end} = n_{**}=2.5\times 10^{14}P_6$ cm$^{-3}$ while $M_{**}\approx 3\times 10^{10}P_6^3$ g and $M_{abl}^{max}\approx 3\times 10^{10}(P_6\sec\phi)^3$ g.
In the case of impacts with Jupiter $H_8 \sim 0.01-0.1$ so, for relatively steep impacts ($\sec \phi \preceq 2$), Eqns. (\ref{NablovrNPres}, \ref{MAblmax}) show that ram pressure will cause explosion before ablation is complete for $M_o\succeq 2\times 10^9P_6^3$ g. Thus, as argued by Chyba et al (1993), Zahnle and MacLow (1994) and Carlson et al. (1997), for steep planetary impacts the ablation process is secondary to pressure-driven explosion except for small masses (or extremely shallow entry) and during the initial entry phase. In the case of solar chromospheric impacts, however, $H_8=0.5$ is larger so that the maximum ablation-dominated mass (after sublimation losses) is around $3\times 10^{10}P_6^3$ g  for steep entry. However, most sun-impactors have $q\approx R_\odot$ and $\sec \phi \gg 10$. Thus the maximum ablation-dominated mass for these is $\gg 10^{14}P_6^3$ g as discussed further in Section 6.1.

In the planetary impact literature two factors are discussed which act to limit ablation and we show briefly here why they are less relevant to the solar case. The first is possible overestimation of the drag coefficient - we have taken the nucleus effective area to equal its geometrical area, which is an overestimation for fluid flow round an obstacle. However, for the solar case the nucleus speed is so high and the atmospheric density so low that we can reasonably use a kinetic description and treat the particles as impacting the nucleus directly rather than flowing around it. We note, however, that our kinetic approach may break down toward the lower chromosphere relevant to very heavy nuclei and steeper incidence angles. There the high density results in a shorter collisional mean free path and more fluid like behavior with only indirect ablation via the radiation from a stand-off shock. This can decrease the drag coefficient considerably so that our ablation and deceleration rates are overestimates there.
Secondly, it has been argued in the planetary case (e.g. Zahnle and MacLow 1994) that ablation is limited by the loss of energy needed to ionize hydrogen. However, the solar atmosphere is around 100 times hotter than Jupiter's atmosphere and is already substantially ionized at least in the upper chromopshere. Further, solar impacting protons have keV kinetic energies - far higher than H ionization energy (13.6 eV) and than the 10 eV value for Jupiter so direct impact ionization is very effective in the sun but not Jupiter. Again, however, heavier and steeper entry nuclei reaching the deep chromosphere
are in a cooler medium of lower ionization so energy lost to ionization may become significant and our ablation rate an overestimate.
\section{Theory of Nucleus Sublimation/Ablation}
\subsection{General Formulation}
Having discussed orders of magnitude we now want to model the evolution of evaporative mass loss along the comet path. We take the nucleus to have constant $\rho$ and to shrink due to mass loss, neglecting for now any pressure driven explosion or distortion which we address later. The changing $M(r)$ and $a(r)$ are then functions of distance $r$ from the solar centre. We treat mass loss (by sublimation or ablation) via the simple approximation that a heating/energy deposition rate of $\cal P$ erg s$^{-1}$ results in mass loss rate $\dot M $ g s$^{-1}= {\cal P}/ {\cal L}$ with the relevant $\rho, {\cal L}$ values averages over the icy conglomerate components (Sections 2.2.1 and 2.2.3). For the sublimation case the dominant power required for loss of mass from the nucleus is that for sublimation of the ices only, the solid dust particles released escaping with the vapor outflow. That is, fairly near the sun at least, the power going into sublimation is larger than that going into further heating offset by cooling - e.g. Weissman (1993), Sekanina (2003). (Far from the sun our neglect of cooling etc will overestimate the mass loss rate). For the ablation case, heating is intense and vaporizes all components, including the most non-volatile ones but (Section 2.3), the overall ${\cal L}$ is similar to ${\cal L}_{ice}$. We further assume that all mass lost clears the nucleus quickly enough for incident energy fluxes to act directly on it - i.e. the mass loss does not obstruct the mass loss driver. To justify this, we note that latent heat $\cal L$ is equivalent to a speed $\sim 2{\cal L}^{1/2}\sim 2.3\times 10^5$ cm/s. If vaporized material outflow occurs at this speed then, on the timescale $H/v_\odot\cos\phi\sim 10$ s relevant to ablation, it clears nuclei of $a_o\preceq 2.3\times 10^6$ or $M_o\preceq10^{19}$ g so our assumption is justified except for very large masses. These are in any case destroyed by explosion rather than ablation. In the case of sublimation the mass loss rate is much lower (Section 3.3) so the approximation is very good.

Then, for an incident energy flux $\cal F$ (erg cm$^{-2}$s$^{-1}$), ${\cal P} \approx a^2 {\cal F}$ and, with $ a^2=(M/\rho)^{2/3}$, and constant $\rho$ and shape, we get

\begin{equation}
\label{dM/dt}
\dot M = \frac{dM}{dt}\approx -\frac{a^2{\cal F}}{{\cal L}}\approx -\frac{M^{2/3}\cal F}{{\cal L} \rho^{2/3}}
\end{equation}
Note that we consider only mass loss in the one hemisphere (cross section $\approx a^2$) containing flux ${\cal F}$ since rotation times are long (Samarachinha et al. 2004) compared to the short near-sun mass-loss time scale.
Using Eqn. (\ref{veloccomps}) to express $dt = -dr/v_{r}$ we can eliminate $t$ from Eqn. (\ref{dM/dt}) and express it as a differential equation for the fractional mass $m=M(r)/M_o$ remaining as a function of $r$, viz.

\begin{equation}
\label{dm/dx}
\frac{1}{m^{2/3}}\frac{dm}{dr}=\frac{R_\odot^{3/2}}{v_\odot}\frac{1}{M_o^{1/3}{\cal L }\rho^{2/3}}\frac{{\cal F}(r)r^{1/2}}{(1-q/r)^{1/2}}
\end{equation}
 This can be integrated for any specified form of ${\cal F}(r)$ with boundary condition $m(r\rightarrow \infty) = 1$.
We have formulated the problem in terms of $r$ dependence since this allows analytic treatment. Conversion to the time domain can be done numerically via Barker's equation (cf. results in Porter 2007, 2008).

\subsection{Heating Energy Flux Terms and regimes}
\begin{itemize}
\item {\bf (a) absorbed insolation energy flux} for albedo $\ll 1$ is
\begin{equation}
\label{Frad}
{{\cal F}_{rad}} \approx f(r)\frac{L_\odot}{ 4\pi r^2}\approx \frac{L_\odot}{ 4\pi r^2}\approx 5\times 10^{10}\left(\frac{R_\odot}{r}\right)^2~{\rm erg~ cm}^{-2}~{\rm s}^{-1}
\end{equation}
where $f(r)$ is the factor by which the heating rate at $r$ exceeds our estimate in Eqn. (\ref{Frad}) where we use the approximation of near radial insolation. As $r\rightarrow R_\odot$ non-radial insolation from the finite solar disk increases and for a sphere just above the photosphere the maximum is $f(R_\odot)=2$ ($=4$ below the photosphere) but this can increase the mass required to survive insolation 8-fold. $f(r)$ declines above $r=R_\odot$,  is shape dependent, and partly offset by the cooling processes we have neglected but our use of $f\sim 1$ must be borne in mind in considering our results.
\item {\bf (b) ablation energy flux} Impacting solar protons have a collisional mean free path (mfp) in nucleus material ($n\approx 10^{23}$ cm$^{-3}$) of order $10^{-6}$~cm (cf. Brown 1972, Emslie 1978) while in the atmosphere at $n>n_*=2.5\times 10^{11}$, the mfp is $\sim 4\times 10^5$ cm, much bigger than most nuclei. Furthermore, the nucleus speed is highly supersonic ($\sim$ Mach 50 in the chromosphere). Thus the bombarding atmospheric protons have neither enough time nor undergo enough collisions to behave as a fluid. We therefore treat the problem as one of surface ablation by this particle kinetic flux rather than as a fluid flow. The particle bombardment energy flux is then
\begin{eqnarray}
\label{Fcoll}
{{\cal F}_{coll}}(r) & = & \frac{1}{2}\rho_\odot(r)v^3(r)
=\frac{1}{2}\mu m_p n(r)v_\odot^3\left(\frac{R_\odot}{r}\right)^{3/2}\nonumber\\
 &= &2.7\times 10^{16} \frac{n(r)}{n_\odot}\left(\frac{R_\odot}{r}\right)^{3/2}~~{\rm erg~cm}^{-2}~s^{-1}
\end{eqnarray}
(cf. meteor mass loss in the earth's atmosphere - e.g. Kaiser 1961, Chapter 8).
Here $n(r)$ is the total (atom + ion) number density (cm$^{-3}$) of hydrogen,
$m_p$ the proton mass and $\mu\approx 1.3$ the mean total mass present (including helium etc) per hydrogen (so that $\rho_\odot = \mu n m_p$) while $n_\odot\approx 10^{17}$  cm$^{-3}$ is the value of $n$ at the photosphere ($r=R_\odot$) which we take as reference point.
\item {\bf ratio of ablation to sublimation energy fluxes} is
\begin{equation}
\label{FcollovrFrad}
\frac{{\cal F}_{coll}(r)}{{\cal F}_{rad}(r)} \approx 4\times 10^5\left(\frac{r}{R_\odot}\right)^{1/2}\frac{n(r)}{n_\odot}\approx  \left(\frac{r}{R_\odot}\right)^{1/2}\frac{n(r)}{n_*}
\end{equation}
where $n_*=n(r_*)\approx 2.5\times 10^{11}{\rm cm}^{-3}$ is the $n$ value at $r=r_*=1.01R_\odot$ where ${\cal F}_{rad}, {\cal F}_{coll}$ are equal. At the photosphere ($n=n_\odot=10^{17}$ cm$^{-3}$), ${\cal F}_{coll}\approx 10^5{\cal F}_{rad}$ which is equivalent to a black body energy flux at $T\succeq 100,000$ K !
\end{itemize}

From Eqn. (\ref{FcollovrFrad}) we see that sublimation dominates over ablation only at distances where $n(r)\prec n_*\approx 2.5 \times 10^{11}$ cm$^{-3}$ where they are equal. This value $n=n_*$ is in the (upper) chromosphere where the density rises very steeply with depth with a fairly constant temperature (of order 10,000 K) and hence mean exponential density scale height $H\approx 5\times 10^7$~cm $\approx 7\times 10^{-4}R_\odot$, viz.
\begin{equation}
\label{densityexponential}
n(r)=n_\odot e^{-(r-R_\odot)/H}
\end{equation}
In this approximation, the values $r=r_*, n=n_*$ occur at a height $z$ above the photosphere of $z=z_*=r_*-R_\odot \approx H\ln(4\times 10^5) \approx 6500$ km $\approx 0.01 R_\odot$. Since $h=H/R_\odot\approx 7\times 10^{-4}$, over a very small $r$ range (a few $H$)
the dominant energy flux changes rapidly from radiative at $z-z_* \gg H$ to collisions at $z-z_* \ll -H$, a $z$ range $\ll R_\odot$. At several $H$ above $z_*$ ablation becomes negligible compared with sublimation so use of analytic expression (\ref{densityexponential}) for $n(r)$ at all heights, results in negligible error in the total energy flux. Over the very narrow range of $1.005 < r/R_\odot <1.015$ around $r=r_*$, ${\cal F}_{rad}(r)\propto r^{-2}$ increases by 1\% while ${\cal F}_{coll}(r)$ falls $\times 10^6$. Throughout the present paper, for simplicity we approximate $H$ by a height-independent mean value though $H$ changes slowly with depth since the temperature and hydrogen ionization vary. The variation in $H$ is around a factor of 4 over the entire range $r_*\rightarrow R_\odot$ (and far less over the much smaller range of $r$ which proves to be important) while $n(r)$ increases by a factor $10^6$. A detailed model of $n(r)$ over a large range of $r$ was developed by Porter (2007, 2008) and should be used in more precise analysis. In addition chromospheric $n$ structure varies on small horizontal scales about the mean stratified form $n(r)$ so that fluctuating mass loss rate is expected along the path.

From the above discussion, the total energy flux ${\cal F}(r)$ is reasonably well approximated by
\begin{eqnarray}
\label{Ftot}
{\cal F}(r) &= &{\cal F}_{rad}(r)+{\cal F}_{coll}(r)\nonumber\\
&= & \frac{L_\odot}{ 4\pi r^2}+ \frac{1}{2}\mu m_p n_\odot v_\odot^3 e^{-(r-R_\odot)/H}\left(\frac{R_\odot}{r}\right)^{3/2}
\end{eqnarray}
If we define dimensionless variables $x=r/R_\odot, m(x)=M(r)/M_o, p=q/R_\odot,  h=H/R_\odot$, and use Eqn. (\ref{Ftot}) for ${\cal F}(r)$, differential Eqn. ({\ref{dM/dt}}) becomes, with (incoming) boundary condition $m(r\rightarrow \infty_-)=1$
\begin{equation}
\label{dmdx}
\frac{1}{m^{2/3}}\frac{dm}{dx}=\alpha\left[\frac{1}{x(x-p)^{1/2}}
+\beta\frac{e^{-(x-1)/h}}{x^{1/2}(x-p)^{1/2}}\right]
\end{equation}
where the dimensionless parameters $\alpha,\beta$ are
\begin{equation}
\label{alphadef}
\alpha =\frac{L_\odot}{4\pi{\cal L}\rho^{2/3}v_\odot R_\odot M_o^{1/3}}=\frac{0.27}{{\cal M}_{12}^{1/3}}\approx \frac{0.43}{ M_{12}^{1/3}}
\end{equation}
where ${\cal M}_{12}={\cal M}_o/10^{12}{\rm g}, M_{12}=M_o/10^{12}{\rm g},$
\begin{equation}
\label{betadef}
\beta= \frac{2\pi\mu m_pR_\odot^2v_\odot^3n_\odot}{L_\odot}=3.8\times 10^5
\end{equation}
and we define an {\it effective mass}
\begin{equation}
\label{effectivemass}
{\cal M} =\tilde{{\cal L}}^3\tilde{\rho}^2 M
\end{equation}
In Figure 1 we show the dependence of $M_o/{\cal M}_o$ on $\tilde{\rho}, \tilde{{\cal L}}$. $\cal M$ is the mass of water ice which would sublimate at the same rate (g/s) as the actual nucleus mass $M$ with its specific values of $\tilde{\rho},\tilde{{\cal L}}$.
Note also, for dimensional clarity, that $M_{12}$ and ${\cal M}_{12}$ in all equations here are dimensionless, the dimensional scaling unit $10^{12}$ g being included as a factor in the (overall dimensionless) numerical value of $\alpha$.
\begin{figure}
 \resizebox{\hsize}{!}
{\includegraphics{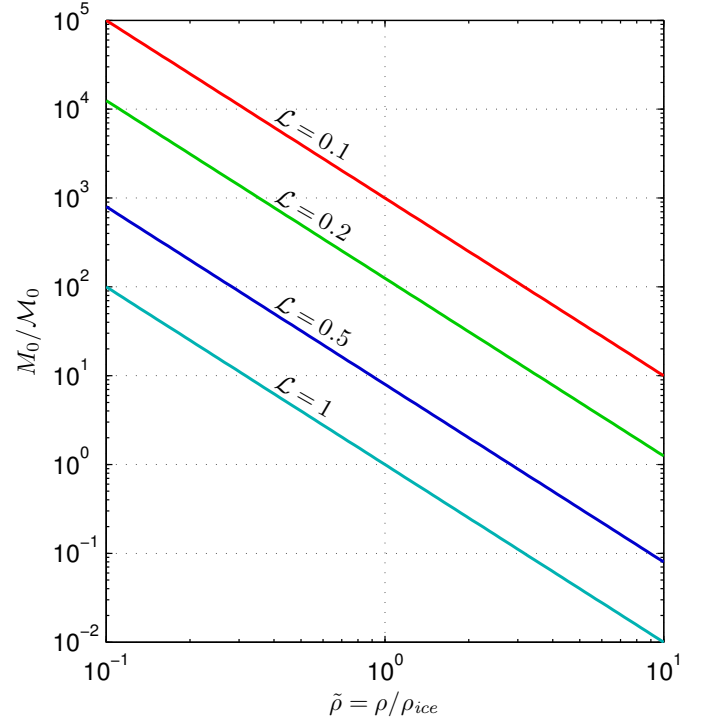}}
\caption{Conversion factor $M_o/{\cal M}_o$ from ice equivalent mass ${\cal M}_o$ to real mass $M_o$ versus $\tilde{\rho}$ for $\tilde{{\cal L}}= 0.1, 0.2, 0.5, 1.0$ - Eqn. (\ref{effectivemass}). For our canonical values $\tilde{\rho}=0.5, \tilde{{\cal L}}=1$, the true mass $M_o=4\times {\cal M}_o$ since comet nuclei are less dense than ice so easier to evaporate.}
\label{fig1}
\end{figure}

In cases where the sole processes acting on the nucleus were sublimation and ablation (ie hydrodynamic explosion effects were small prior to total vaporisation) the {\it pure sublimation/ablation solution} $m(x)=M(r)/M_o$ of Eqn. (\ref{dmdx}) is, on the inbound trajectory ($\infty_->x>p$),

\begin{eqnarray}
\label{m-(x)general}
& & m(x)= m_-(x)\nonumber =\\&=& \left[1-\frac{\alpha}{3}
\left(\int_x^\infty\frac{dx}{x(x-p)^{1/2}}+
\beta\int_x^\infty\frac{e^{-(x-1)/h}dx}{x^{1/2}(x-p)^{1/2}}\right)\right]^3\nonumber\\
&=& \left[1-\frac{\alpha}{3}
\left(\frac{2}{p^{1/2}}\sin^{-1}\left(\frac{p}{x}\right)^{1/2}+
\beta\int_x^\infty\frac{e^{-(x-1)/h}dx}{x^{1/2}(x-p)^{1/2}}\right)\right]^3
\end{eqnarray}
while on the outbound trajectory ($q<x<\infty_+$) it is (with $\int..dx = \int_q^\infty..dx+\int_q^x..dx$)

\begin{eqnarray}
\label{m+(x)general}
& & m(x)= m_+(x)= \nonumber \\
& & \Biggl[1-\frac{\alpha}{3}\times
\Biggl(\frac{2}{p^{1/2}}\times
\big[\pi-\sin^{-1}\Bigl(\frac{p}{x}\Bigr)^{1/2}\big] \nonumber\\
& & + \beta \times \Bigl(2\int_p^\infty\frac{e^{-(x-1)/h}dx}{x^{1/2}(x-p)^{1/2}}-
\int_x^\infty\frac{e^{-(x-1)/h}dx}{x^{1/2}(x-p)^{1/2}}\Big)\Biggr)\Biggr]^3
\end{eqnarray}

Since $L_\odot, R_\odot, v_\odot, n_\odot, H$ are all known, for specified $\rho, {\cal L}, P_c$, this solution $m(x)$ depends solely on the the parameters $q$ and ${\cal M}_o$ and we will see below that the combination ${\cal M}_oq^{3/2}$ appears frequently in the solution. Eqns. (\ref{m-(x)general},\ref{m+(x)general}) for $m(x)$ allow derivation of any property of $M(r)$. For example the mass loss rate $\dot{M}= M_o m'(x)\dot{x}$ where $\dot{x}=\dot{r}/R_\odot=(v_\odot/R_\odot)(1/p/x)^{1/2}/x^{1/2}$ by Eqns. (\ref{veloccomps}). Very near the sun $x\rightarrow 1$ and is not a convenient variable since $1-p/x\rightarrow 0, m'(x)\rightarrow \infty$ and rates should be expressed via the $\theta$ coordinate with $\dot{\theta}=(v_\odot/R_\odot)(p^{1/2}/x^2)$.

From the discussion above it is apparent that comets with $q-R_\odot \succeq 0.01R_\odot$ never enter the ablation regime so can be treated neglecting the ablation terms. On the other hand comets with $q-R_\odot \preceq 0.01R_\odot$, and of large enough ${\cal M}_o$ to survive sublimation, encounter the ablation regime. If, further, they are of large enough ${\cal M}_o$ to survive to $r=r_* = 1.01R_\odot$ with only a {\it small} mass loss fraction $\Delta m=1-m(x_*=r_*/R_\odot) \ll 1$ then their $m(x)$ can be well approximated by including only the ablation terms and neglecting the preceding sublimation $\Delta m$. These sublimation and ablation dominated regimes are treated analytically in Sections 5 and 6.1 while the role of ram pressure driven explosion, vital in the destruction of larger masses or steep entry angle, is discussed in Section 6.2.

\section{Sublimation Dominated Mass Loss Solutions}(${\cal M}_o \preceq 10^{11}$ g or $q-r_*\gg H \approx 0.01R_\odot$)

\subsection{Sublimation dominated solution for m(x)}
In this case solution (\ref{m-(x)general}, \ref{m+(x)general}) for $m(x)$ along the incident and outgoing paths become
\begin{eqnarray}
\label{m(x)rad}
m^{rad}_-(x)&=&\left[1-\frac{2\alpha}{3p^{1/2}}\sin^{-1}\left(\frac{p}{x}\right)^{1/2}\right]^3\nonumber\\
&\rightarrow &\left[1-\frac{2\alpha}{3x^{1/2}} \right]^3~{\rm as}~ p\rightarrow 0 \nonumber \\
m^{rad}_+(x)&=& \left[1-\frac{2\alpha}{3p^{1/2}}
\left(\pi - \sin^{-1}\left(\frac{p}{x}\right)^{1/2}\right)\right]^3
\end{eqnarray}
which is plotted in Figure 2 in terms of $m(x/p)$ for various values of the parameter ${\cal M}_op^{3/2}$. In the following subsections we discuss features of this solution which are of particular interest.
\begin{figure}
 \resizebox{\hsize}{!}
{\includegraphics{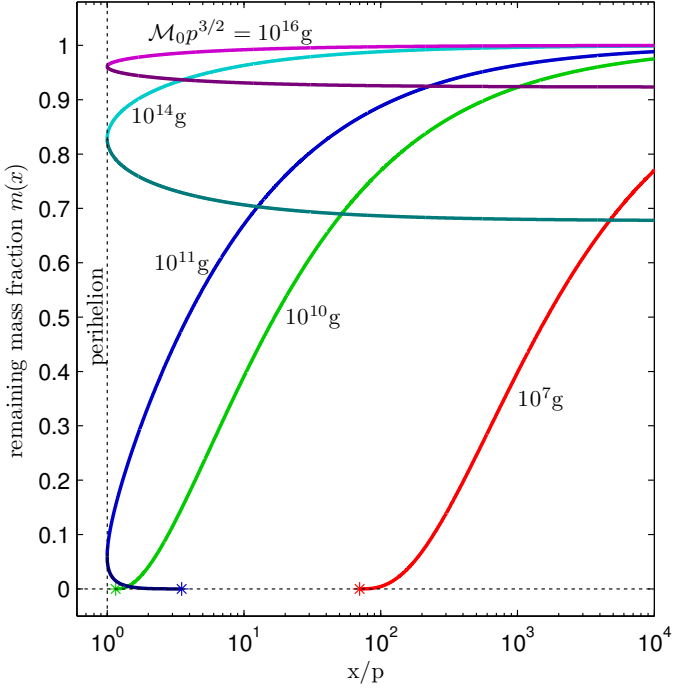}}
\caption{$m(x)$ versus $x/p$ for various ${\cal M}_op^{3/2}$ showing cases of survival of sublimation through an orbit, and of terminal mass loss before and after
perihelion. This Figure can be used for any values of $M_o, p,\rho, {\cal L}$ because of the scaling and combination of variables used.}
\label{fig2}
\end{figure}

\subsection{Sublimation solution properties}
\subsubsection{Fractional mass lost by perihelion/after one orbit}
By Eqn. (\ref{m(x)rad}) with $x=p=q/R_\odot$,
 the fractional masses $m$ surviving, and $\Delta m$ lost by sublimation to perihelion are
\begin{eqnarray}
\label{mq}
m^{rad}_q &=1-\Delta m^{rad}_q =\left[1-\frac{\pi\alpha}{3p^{1/2}}\right]^3& \nonumber \\ &\approx\left[1-\frac{0.27}{p^{1/2}{\cal M}_{12}^{1/3}}\right]^3
\approx\left[1-\frac{0.43}{p^{1/2}M_{12}^{1/3}}\right]^3&
\end{eqnarray}
while after one orbit they are,  by Eqn. (\ref{m(x)rad}) for $m_+(x\rightarrow \infty$)
\begin{eqnarray}
\label{morbit}
m^{rad}_{orbit}=1-\Delta m^{rad}_{orbit} = \left[1-\frac
{2\pi\alpha}{3p^{1/2}}\right]^3 \nonumber \\ \approx\left[1-\frac{0.56}{p^{1/2}{\cal M}_{12}^{1/3}}\right]^3\approx\left[1-\frac{0.86}{p^{1/2}M_{12}^{1/3}}\right]^3
\end{eqnarray}
the last forms being for our canonical $\tilde{\rho},\tilde{{\cal L}}$ values. (These expressions are valid for $p>\pi^2\alpha^2/9,4\pi^2\alpha^2/9$ respectively). Note that $\Delta m = \Delta M/M_o=\Delta {\cal M}/{\cal M}_o$, and that the minimum initial mass needed to just survive sublimation through one orbit is 8 times that needed to just survive to perihelion (the corresponding minimum $a_o$  values being in the ratio 2:1). We can rewrite Eqn. (\ref{mq}) as
\begin{equation}
\label{MovsDeltamq}
p^{3/2}{\cal M}_{12}=\frac{0.022}{[1-(1-\Delta m_q^{rad})^{1/3}]^3}
\end{equation}
which lets us evaluate the initial ${\cal M}_o\approx 0.25M_o$ which will lose a fraction $\Delta m$ by perihelion - see Figure 3.

\begin{figure}
 \resizebox{\hsize}{!}
{\includegraphics{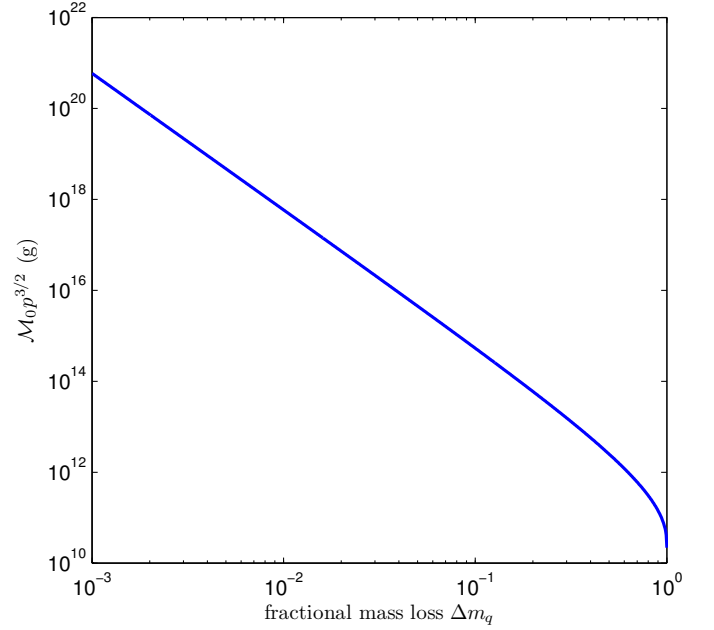}}
\caption {Value of ${\cal M}_op^{3/2}$ which has undergone fractional sublimation mass loss $\Delta m_q$ by perihelion - Eqn. (\ref{MovsDeltamq}). The Figure can be used for any values of $Mo,p,\rho, {\cal L}$ because of the scaling and combination of variables used}
\label{fig3}
\end{figure}

\subsubsection{Minimum initial masses and radii surviving sublimation to perihelion and through one orbit}
Eqns. (\ref{mq}, \ref{morbit}) imply the following for the minimum initial masses needed to just survive sublimation to perihelion and through one complete orbit.
\begin{eqnarray}
\label{Mominq}
M^{rad}_{ominq}(q)&= \frac{{\cal M}_{ominq}^{rad}}{\tilde{{\cal L}}^3\tilde{\rho}^2}=\frac{M^{rad}_{minorbit}(q)}{8} =\left[\frac{L_\odot}{12{\cal L}\rho^{2/3}v_\odot R_\odot p^{1/2}}\right]^3 & \nonumber \\
&\approx \frac{2.2\times 10^{10} {\rm g}}{\tilde{{\cal L}}^3\tilde{\rho}^2p^{3/2}}\approx \frac{8.8\times 10^{10} {\rm g}}{p^{3/2}}&
\end{eqnarray}
the corresponding minimum initial sizes being
\begin{equation}
\label{aomin}
a_{ominorbit}^{rad}=2a_{ominq}^{rad}\approx \frac{2.7\times 10^3~{\rm cm}}{\tilde{{\cal L}}\tilde{\rho}p^{1/2}}\approx \frac{5.4\times 10^3~{\rm cm}}{p^{1/2}}
\end{equation}
According to Section 2.4 for the case of a sphere this corresponds to radius $b_{ominorbit}\approx a_{ominorbit}/1.67\approx 3.2\times 10^3/p^{1/2}$ cm.
Consequently (for $\tilde{\rho}=0.5,\tilde{{\cal L}}=1$) any $M_o\preceq 10^{11}$ g with $p\approx 1$ will be totally sublimated before reaching the ablation zone while $M_o\succeq 10^{11}$ g with $p\preceq 1.01$ will undergo ablation and be totally dominated by it or by explosion, according to the conditions discussed in Sections 3.4 and 6.1-6.2.

Eqns. (\ref{Mominq}, \ref{aomin}) can be generalised to find the mass ${\cal M}_o$ which is totally sublimated at any point $x_+$ or $x_-$ by setting $m=0$ and solving for ${\cal M}_o$. Thus observation of such $x_{end}$ values allows nucleus mass estimation as indeed does any measurement of mass loss or mass loss rate at specific $x$. Note that specifying $m$ at $x$ does not define ${\cal M}_o$ uniquely unless one know if it is $x_+$ or $x_-$. A small ${\cal M}_o$ has the same $m$ at inbound point $x$ as a corresponding large ${\cal M}_o$ has when outbound at $x$.

From all of the above it is apparent (cf. Section 3.3) that, near the sun, mass loss by sublimation, and the subsequent deposition of kinetic energy and abundance-anomalous matter in the corona occur over distance scales of order $R_\odot$ and timescales of order $R_\odot/v_\odot$. So even total destruction of the nucleus is a slow extended 'fizzle', not an explosion (cf observations by Schrijver 2011 for a case with $p\sim 1.14$). This contrasts sharply with low altitude ablation/explosion over a few scale heights $H\approx 10^{-3}R_\odot$ directly powered by nucleus kinetic energy - Section 6.
\subsection{Comparison with Numerical Models}
In order to see whether our simple analytic expressions above give reasonable approximations to mass loss, at least for $q\preceq$ a few $R_\odot$, we can compare them with those from numerical models incorporating effects we have neglected, such as the early numerical thermal modeling by Weissman and Keiffer (1982) and Weissman (1983). Weissman (1983) (his Fig. 2) predicts for water ice spheres the change $\Delta a$ resulting from sublimation through infall to $q=R_\odot$ to be $\Delta a\approx 2\times 10^3$ cm for zero albedo. This can be compared with our expression (\ref{aomin}) with $p=1$ namely, (with $\tilde{\rho}=1, \tilde{{\cal L}}= 1$ for water ice)  $a_{omin}\approx 2.7\times 10^3$ cm. Converting this to the equivalent radius for the case of a sphere (Section 2.4)
gives $1.6\times 10^3$ cm in good (20\%) agreement with Weissman, given our approximations (e.g. radial insolation and no cooling processes). This gives us confidence in applying our analytic approach to get estimates for the ablation regime (Section 6).

\section{Ablation and Explosion of Sun-impactors}
\subsection{Ablation Dominated Solution when ram pressure can be neglected - $M_o(1-q/r_*)^{3/2}\preceq M_{**}\approx 10^{10}{\rm g}, r_*-q\succeq H$}
For nuclei with $q\preceq r_*$ and with original $M_o$ large enough $\succeq 10^{11}$ g to survive sublimation down to $r<r_*$ the residual mass $M'_o=M_o(1-\Delta m(x_*))$ has (cf. Section 5.2.1) essentially the same velocity as if it had arrived at $r_*$ without any sublimation. Because of the very rapid large switch from the sublimation to the ablation regime, its subsequent mass loss can thus be treated using only the ablation term in our $m(x)$ expression (\ref{m-(x)general}) for a mass $M'_o$ falling from infinity. For simplicity here we will only consider $M_o$ large enough that $\Delta m(x_*)\ll1$ and simply equate $M'_o=M_o$. Consequently our results for masses in the transition range around $10^{11}$ g are approximate. In this Section we further restrict ourselves to impactors having small enough $M_o$ and/or shallow enough incidence that atmospheric ablative mass loss is complete before ram pressure driven explosion is significant (exploding impactors are discussed in Section 6.2). The inbound evolution of $m(x)=M(r)/M_o$ by ablation alone can then be well described by using only the ablation terms in Eqn. (\ref{m-(x)general}). That is
\begin{equation}
\label{mcoll-(x)}
m^{coll}_-(x)=m_{coll}(x)=\left[1-\frac{\alpha\beta}{3}
\int_x^\infty\frac{e^{-(x-1)/h}dx}{x^{1/2}(x-p)^{1/2}}\right]^3
\end{equation}

The integral in Eqn. (\ref{mcoll-(x)}) has in general no standard closed analytic form but, since $h\ll 1$, a good analytic approximation is readily obtained for almost all $p$. Expression (\ref{mcoll-(x)}) is only physically relevant down to where ablation is complete - viz $\int.. dx \le 3/\alpha\beta$ - which always occurs close to $x=x_*\approx 1.01$ since $h\ll 1$. Below $x=x_*$ the integrand increases exponentially with depth so is maximal at the $x$ limit of the integral. Consequently, (unlike the exponential numerator) the integrand denominator can be well approximated by setting $x_*(x_*-p)^{1/2}\approx (1-p/x_*)^{1/2}$ (Rare cases where $|x_*-p| \preceq h$ need a more refined treatment, and inclusion of both ablation and sublimation). Solution  (\ref{mcoll-(x)}) can then be integrated to give

\begin{equation}
\label{m(x)coll}
m(x)=m_{coll}(x)\approx \left[1-\Gamma \frac{e^{-(x-1)/h}}{(1-p/x_*)^{1/2}}\right]^3
\end{equation}
where
\begin{equation}
\label{Gammadef}
\Gamma = \frac{\alpha\beta h}{3}=\frac{\mu n_\odot m_pv_\odot^2H}{6{\cal L}\rho^{2/3}M_o^{1/3}}\approx \frac{30}{({\cal M}_o/10^{12})^{1/3}}
\end{equation}
from which it follows that the nucleus will be totally ablated by the height
\begin{eqnarray}
\label{zablend}
&&z_{abl}^{end}({\rm km})= H\ln \frac{\Gamma}{(1-p/x_*)^{1/2}} =\nonumber \\&&  2000
+1100\times\left[\frac{1}{2}\log_{10}(1-p/x_*)-\frac{1}{3}\log_{10} M_{12}\right]
\end{eqnarray}
with corresponding atmospheric density
\begin{eqnarray}
\label{nablend}
&&n_{abl}^{end}({\rm cm}^{-3})=N_{abl}^{end}/H =\frac{n_\odot(1-p/x_*)^{1/2}}{\Gamma} \nonumber \\
&&\approx 1.25\times 10^{15}(1-p/x_*)^{1/2}M_{12}^{1/3}
\end{eqnarray}

Eqn. (\ref{zablend}) shows that, even neglecting explosion, all nuclei are destroyed over a few scale heights vertically. Such short paths suggest that the ablation path may be approximated as rectilinear, starting around the point where $n\approx n_*$ at angle $\phi\approx \phi_*$ to the vertical, and of nearly constant speed $v_\odot$. From Eqns. (\ref{veloccomps}) we find that

\begin{equation}
\label{cosphistar}
\cos\phi_* =v_r(r_*)/v(r_*)=(1-p/x_*)^{1/2}
\end{equation}

If one re-derives and solves the equation for $dm/dx$ for a rectilinear path one arrives at exactly the same expressions as
found in Eqns. (\ref{zablend}, \ref{nablend}) given the relation (\ref{cosphistar}) of $\phi_*$ to $p/x_*$ so our algebraic approximation there is equivalent to the geometric one of a rectilinear path.

\begin{figure}
\resizebox{\hsize}{!}
{\includegraphics{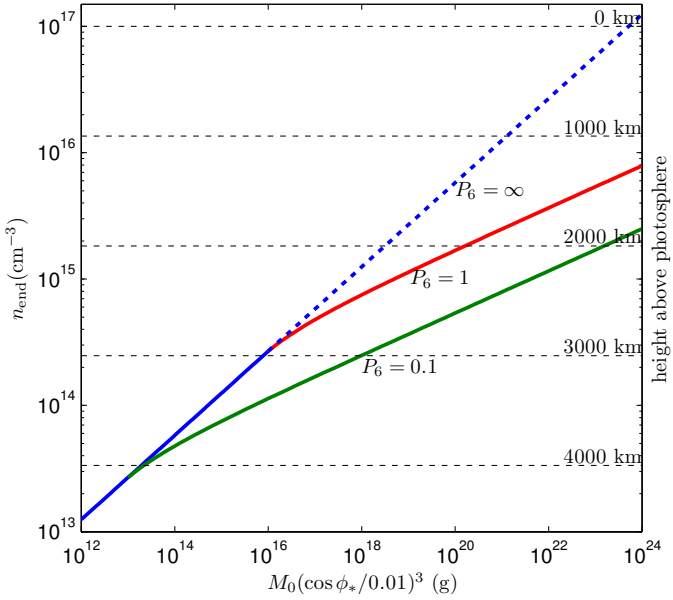}}                                                          \caption{Estimate of the heights $z_{end}$ and densities $n_{end}$ at which nuclei are destroyed as a function of $M_o \cos^3\phi_*$ for specified $P_6$. Note that results are very sensitive to $\cos\phi_*$ and to $P_6$. To the left of the knee ablation vaporizes the nucleus before ram pressure exceeds the strength of nucleus material. To the right ram pressure drives destructive explosion of the nucleus. It is clear that a very large range of masses is destroyed within a few scale heights in the chromosphere and that for all plausible masses and parameters, all nuclei are destroyed above the photosphere.}
\label{f4}
\end{figure}

From Eqn. (\ref{NablovrNPres}) it then follows that ram pressure driven explosion is small throughout the total ablation phase ($n_{abl}^{end} \preceq n_{**}=2.5\times 10^{14}P_6$) cm$^{-3}$ only for masses smaller than

\begin{eqnarray}
\label{Momaxofphi}
M_o & \preceq M_{abl}^{max}(\phi)= \left[\frac{P_cH}{\tilde{{\cal L}} \tilde{\rho}^{2/3}}\right]^3(1-p/x_*)^{-3/2}=M_{**}\sec^3\phi_*=& \nonumber \\
&\approx \frac{ 10^{10}P_6^3}{(1-p/x_*)^{3/2}}\approx 10^{10}P_6^3 \sec^3\phi_* ~~{\rm g}&
\end{eqnarray}
 where we have used $H_8\approx 0.5$ and $\tilde{{\cal L}}=1, \tilde{\rho}= 0.5$. Here $M_{**}$ is the maximum vertical entry mass which is totally ablated by depth $n=n_{**}$. These values of $n_{abl}^{end}, M_{abl}^{max}$ differ by factors of a few from the estimates in Section 3.4.2 due to the more accurate treatment of ablation. Since $x_*=1.01$, for the range $1<p<1.0099$, the final $\sec^3\phi_*$ factor is in the range $10^3-10^6$ and the mass above which explosion becomes dominant for such grazing impactors is in the range $3\times 10^{13-16}$ g for $P_6=1$ and $3\times 10^{10-13}$ g for $P_6=0.1$. This solar limit is much larger than for the conditions of the Shoemaker-Levy 9 Jupiter impacts (with fragment masses estimated to be around $10^{15}$ g) because of Jupiter's smaller $H$ and steep angle of those impacts ($\sim 45\deg $) which result in a much larger density $n$ and hence ram pressure at a specified ${\cal N}$ (Section 3.4.2). This modifies somewhat the relative importance and regimes of ablation and explosion in nucleus destruction in the solar case, as we discuss below after discussing the explosion regime.

 \subsection{Explosive end point for nuclei of high $M_o\cos^{3/2}\phi_*$}
For nuclei of $M_o>M_{abl}^{max}=M_{**}\sec^3\phi_*$, mass loss and destruction of the nucleus become dominated by expansion due to the action of the exponentially growing ram pressure. This is compounded by breakup of the nucleus and flow via hydrodynamic instabilities -  see  Chyba et al (1993), Field and Ferrara (1994), MacLow and Zahnle (1994), Zahnle and MacLow (1994) and Carlson et al (1997).
Modeling this intense localised explosion and the subsequent fireball expansion in the solar case is beyond the scope of the present exploratory paper and should be addressed using modified versions of the numerical simulations conducted for planetary cases. Here we simply make an estimate of how much the end depth is reduced by explosion and by the enhanced deceleration and ablation arising from the growing area, based on the analytic approximation given by the above authors.
The explosive lateral expansion of nuclei can be approximated as
incompressible longitudinal shortening combined with lateral expansion
with lateral size $a_\perp$ increasing as (MacLow and zahnle 1994
Equation (2))
 \begin{equation}
 \label{explradius}
 a_\perp(N)= a'_{o}\left(\frac{N}{N_{**}}\right)^{1/2}
 \end{equation}
where $a'_{o}$ is the size when pure ablation ends at $N=N_{**}$ given by (Equation (\ref{m(x)coll}))
 \begin{equation}
 \label{aone}
 \frac{a'_{o}}{a_o}= \left(\frac{M'_{o}}{M_o}\right)^{1/3}= 1-\frac{N_{**}}{N_{abl}^{end}(M_o)}
 \end{equation}
 with $N_{abl}^{end}(M_o)$ (Equation (\ref{nablend})) the depth (hypothetical for $M>M_{abl}^{max}$) at which ablation would be complete in the absence of explosion.
 Determination of the final destruction depth in the explosion regime requires numerical simulation but as an upper limit we can evaluate the depth at which ablation across the exploding area alone would destroy the nucleus. The relevant rate is
  \begin{equation}
 \label{dMdNexp}
 \frac{dM}{dN}=-\frac{\mu m_p v_\odot^2\sec\phi (a'_{o})^2}{6{\cal L}}\frac{N}{N_{**}}
 \end{equation}
 with solution
  \begin{equation}
 \label{Mexp(N)}
 M(N)=M'_{o}\left[1-\frac{\mu m_p v_\odot^2\sec\phi (a'_{o})^2}{12{\cal L}M_{**}} \left( \left(\frac{N}{N_{**}}\right)^2-1\right)\right]
 \end{equation}
 with resulting end depth ($N_{end}^{exp}=Hn_{end}^{exp}$)
 \begin{eqnarray}
 \label{Nexpend}
 && N_{end}^{exp}=N_{**}\left[1+\frac{12{\cal L}\rho^{2/3}\cos\phi M_o^{1/3}}{\mu m_p v_\odot^2N_{**}}\left(1-\frac{N_{**}}{N_{end}^{abl}(M_o)}\right)\right]^{1/2}\nonumber\\
 &&=N_{**}\left[\frac{N_{end}^{abl}(M_o)}{N_{**}}-1\right]^{1/2}~~,~~ M_o \succeq M_{abl}^{max}
 \end{eqnarray}

Combining Equations (\ref{nablend}) and (\ref{Nexpend}) we show in Figure 4 the estimated density $n_{end}=N_{end}/H$ and height $z_{end}=H\ln(n_\odot/n)$ at which destruction ends as a function of $M_o(\cos\phi/0.01)^3$ for values of $P_6=0.1,1$. $P_\infty$ is also shown to make clear how explosion shortens the path relative to ablation alone. The x-axis values are the actual $M_o$ values for $\cos\phi=0.01$ which maximises the ablation regime. For photospheric $p=1, \cos\phi=0.1$ the mass values are $1000$ times smaller. Thus as one decreases $P_6$ and as $p/x_*$ becomes progressively smaller than unity, the mass regime of ablation dominated destruction becomes smaller and smaller. We conclude that for grazing solar impacts, for $P_c\succeq10^5$ dyne/cm$^2$ our ablation-dominated treatment is appropriate for masses well above the sublimation survival limit only when the entry angle is very shallow (i.e. $q$ is very near to $r_*=1.01R_\odot$). For larger $M_o$ or lower $P_c$ or  lower $p$ (less shallow entry), ram pressure driven explosion of the nucleus becomes wholly dominant. For solid objects with $P_6 \gg 1$ the ablation regime is much larger.

\subsection{Energy deposition profiles of Sun-impacting comets}
To calculate observable diagnostic signatures of the destruction of comets near the sun requires modeling the primary energy deposition profile then calculating the thermal, hydrodynamic, and radiative response of the solar atmosphere - the evolution of the fireball which follows the primary airburst. For the impact case, doing this numerically along somewhat similar lines to planetary impact work by e.g. Carlson et al. (1995, 1997) and other work cited in Sections 3.4.1 and 6.2 should be done but is beyond the scope of the present paper. Here we simply take a first step by estimating the height distribution of the nucleus kinetic energy deposition for the ablation dominated case, touching briefly on the explosion domain.

The height distribution $d{\cal E}/dr$ (erg/cm) of kinetic energy deposited along the ablation path can be approximated by
\begin{equation}
\label{heatingperunitht}
\left[\frac{d{\cal E}}{dr}\right]_{abl}\approx \frac{1}{2}v^2(r)M_o\frac{dm}{dr}=\frac{M_ov_\odot^2}{2R_\odot x}\frac{dm}{dx}\approx \frac{M_ov_\odot^2}{2R_\odot }\frac{dm}{dx}
\end{equation}
since $x$ is very close to unity and $v$ very close to $v_\odot$ over the few scale heights involved.

Using Eqn. (\ref{m(x)coll}) to find $dm/dx$ leads to the conclusion that
$d{\cal E}/dr$ is very sharply peaked in height $z$ ($\Delta z\approx H$) with a maximum of
\begin{equation}
\label{dEcaldrmax}
\left(\frac{d{\cal E}}{dr}\right)_{max}\approx 2\times 10^{19}(M_o/10^{12})\sec\phi_*~{\rm erg/cm}
\end{equation}
at
\begin{equation}
\label{nmax}
n_{max}({\rm cm}^{-3})\approx 2\times 10^{14}(1-p/x_*)^{1/2}( M_o/10^{12})^{1/3}
\end{equation}

In the rare cases of steep entry/higher masses we have seen in Section 6.2 that the onset of explosion causes rapid deceleration and disintegration of the nucleus. This will result in even higher volumetric heating rates than in the case of pure ablation, though numerical simulations are needed to quantify this. Such atmospheric power input is similar to that in solar magnetic flares ($10^{27-30}$ erg/sec over loop lengths $\sim 10^9$ cm).
\section{Discussion and Conclusions}
\subsection{Summary of Nucleus Destruction Regimes}
We have shown there to be three regimes of nucleus destruction, depending on the values of $M_o,q$ for given values of $\tilde{\rho}, \tilde{{\cal L}}$ (for which we use here $\tilde{\rho}=0.5, \tilde{{\cal L}}=1$), namely
\begin{itemize}
\item {\bf (a) Sublimation dominated} (Section 5.2.2, Eqn. (\ref{Mominq}))

Nuclei of $p \succeq x_*, M_o \preceq M_{**}(p)$ given by
\begin{equation}
\label{M1ofp}
M_1(p) =M_{ominq}^{rad}=\frac{M_{*}}{p^{3/2}}\approx \frac{10^{11}}{p^{3/2}}~{\rm g}
\end{equation}
are completely sublimated before they reach perihelion, (or $r_*$).
This is by far the commonest regime for group comets.

\item {\bf (b) Ablation dominated} (Sections 3.4.2, 6.1)

Since $n=n_{**}=2.5\times 10^{14}P_6$ at $x_{**}$,
where atmospheric ram pressure starts to exceeds nucleus strength,
nuclei with $x_{**}<p<x_*$ and $M_1(p)<M_o<M_2(p)$ (i.e. of sufficiently
low mass or shallow entry angle) lose all their mass to ablation before
explosion with

\begin{eqnarray}
\label{M2ofp}
&&M_2(p) = M_{abl}^{max}(p)=\nonumber \\
&&M_{**}P_c^3(1-p/x*)^{-3/2}\approx 10^{10}P_6^3(1-p/x*)^{-3/2}{\rm g}
\end{eqnarray}
Since most larger sun-grazers have $p>x_*$, these are rarer than comets in the sublimation regime.

\item {\bf (c) Explosion dominated} (Section 6.2)

Nuclei with $p<x_{**}$ and $M_o>M_2(p)$
(ie of sufficiently high mass or steep entry) undergo ram pressure driven
explosion. These are also rare compared to sublimative destruction.
\end{itemize}
These domains in the plane of $(p/x_*,M_o)$  are summarized Figure 5 with the $M_o$ axis scaled with respect $P_6^3$. $P_6=1$ represents relatively strongly bound nuclei. For $P_c\sim 10^5$ dyne/cm$^2$ the ablation regime becomes confined to about $M_o\succeq 10^{11} {\rm g}, 0.99\preceq p/x_*\preceq 1$. Though this is very narrow in $p$ space,
 how common ablation domination is compared to explosion domination depends on the distribution of $p/x_{**}$ and is very sensitive to the value of $P_c$. For very weakly bound bodies ($P_6 \ll 0.1$) ablation would only dominate briefly prior to explosion totally dominating. Solid stony or iron bodies with $P_6\succeq 10$ would, unless extremely massive, be wholly in the ablative destruction regime.

 \begin{figure}
 \resizebox{\hsize}{!}
{\includegraphics{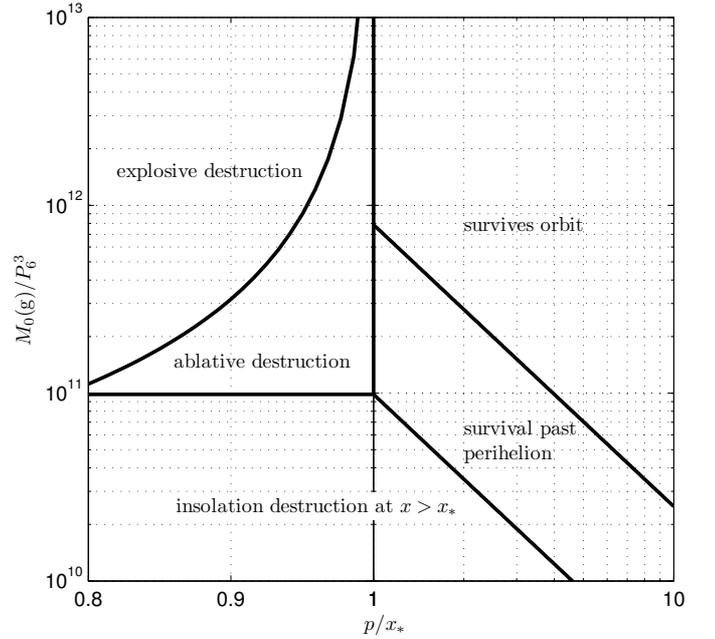}}
\caption{Domains of destruction of comet nuclei in the plane ($M_o/P_c^3,p/x_*=\sin \phi_*$) for entry angle $\phi_*=\arcsin(p/x_*)$ to the vertical at $r=r_*$. NOTE the change from linear to log scale on the $p/x_*$ axis across the point $p/x_*=1$ so as to display the very rapid variation in the range $p/x_*\le 1$ The axes are
scaled with respect to $P_6$ as discussed in section 2.2. For $P_6=0.1$ the ablation regime becomes a slender strip at $0.99\preceq p/x_*\preceq1$}
\label{fig5}
\end{figure}
\subsection{Visibility of Sun-grazer destruction by sublimation}
It was shown in Section 3.3 that destruction by sublimation of low mass nuclei even very near the sun occurs over scales $\sim R_\odot$ in space and $\sim 10^{3-4}$ s in time, making them hard to observe against solar atmospheric emission and scattering. However, the abrupt collisional stopping of sublimated material in the wake could raise its initial temperature by up to around $10^7$ K which may make it visible in the XUV by charge exchange or thermal line emission if it does not cool too quickly by radiation, conduction or expansion and if its emission measure $\int_Vn^2dV$ is high enough. (The first observation, in XUV, of such
an event was made
by Schrijver et al., 2011, using SDO AIA during the revision stage of this paper). A $10^{12}$ g mass totally sublimated over a $0.5 R_\odot$ path into a conical wake of half angle say $10^{-2}$ radians would have an emission measure of order $3\times 10^{44}$ cm$^{-3}$ and a mean density of $\sim 5\times 10^8$ cm$^{-3}$ in a solar plasma environment of roughly similar density.

\subsection{Signatures of Sun-impactor Destruction}
\subsubsection{Mass, Energy and Temperature of the Fireball}
The kinetic energy initially released in the lower solar atmosphere by sun impacting cometary nuclei with $M_o=10^{11}-10^{19}$ g  is $5\times (10^{26}-10^{34})$ erg, similar to the range of the magnetic energy release from 'nano'-flares to 'super'-flares. If 50\% of ${\cal E}_{kin}$ went into heating the comet material the temperature attainable would be $T>m_pv_\odot^2/4k\approx 10^7 K$ where $k$ is Boltzmann's constant. This is of order the solar escape temperature since it is generated by infall of matter effectively from $\infty$.

Flare masses ejected are of the order of $10^{12-16}$ g,  also comparable with those of larger impacting cometary nuclei. Thus a comet impact in the dense chromosphere should produce phenomena somewhat resembling a solar (magnetic) flare, though the initial comet volume is much smaller, its density much higher, and the heating rise time faster ($\preceq 10$ s). Because of the initially small size scale and high density and pressure we might expect rapid initial cooling (seconds) by radiation, expansion and conduction, followed by a radiative/conductive decay phase more like that of solar flares ($\sim 10^{2-3}$ s) as the fireball expands outward and upward, sweeping solar plasma with it (cf. the Carlson et al. 1997 study of the Jupiter impact fireball).

\subsubsection{Cometary Flare Size, Emission Measure and Optical Depth}

If optically thin, the radiative output of a hot plasma of volume $V$ depends on its temperature and its emission measure $EM= \int_Vn^2dV\approx a^3(M_o/\mu_c m_p)^2\approx 10^{48} (M_o/10^{12})^2/(a/10^8)^3/\mu_c$ cm$^{-3}$ for a nucleus of mass $M_o$ when it has expanded to size $a$. (Here $\mu_c$ is the mean comet mass per particle in units of $m_p$). Explosion of a nucleus occurs over a few scale heights $\approx 10^8$ cm vertically, traversed in $\sim 10$ s at speed $v_\odot\cos \phi$. If the lateral expansion speed of the hot wake were similar then the lateral dimension of the exploding wake would also be a few scale heights. Its optical depth for a process with absorption cross-section $\sigma$ would then be $\tau \approx \sigma M_o/m_pa^2 = (M_o/10^{12})(\sigma/10^{-20})/\mu_c/(a/10^8)^2$. Thus when a nucleus of, say, $M_o=10^{12}$ g ($a_o\sim 10^4$ cm) becomes optically thin ($a=10^6$ cm) for $\sigma=10^{-20}$ cm$^2$, it would have an $EM\approx 10^{48}$ cm$^{-3}$ for that emission. This $EM$ is only for the cometary material itself, but the $EM$ of the heated atmosphere may also be important since an atmospheric mass $\succeq M_o$ is involved in decelerating it. Since $H\sim$1 arcsec we may expect to see initially a source of a few arcsec expanding at about 0.1 arcsec per sec at chromospheric altitudes and atmospheric density $\simeq 10^{13-16}$ cm$^{-3}$ depending on $M_o$ as per Figure 4. XUV line spectra should exhibit highly non-solar abundances (eg O:H ratio) as should any matter ejected into space (cf. Iseli et al 2002). There may also be a variety of nonthermal radio and other signatures (e.g. charge exchange - e.g. Lisse et al. 1996 - and impact polarized spectrum lines - e.g. Fletcher and Brown 1992,
1995) arising from plasma phenomena driven by the highly supersonic expansion, such as shock and/or turbulent acceleration of electrons and ions and charge separation currents as the ablated matter decelerates in the atmosphere - cf. review by P\'{e}gauri\'{e}r (2007) of fusion pellet ablation physics.

\subsubsection{Cometary Sunquakes and other solar disturbances}

The phenomenon of flare induced sunquakes - waves in the photosphere -  discovered by Kosovichev and  Zharkova (1998) and now widely studied (e.g. Kosovichev 2006) should also result from the momentum impulse delivered by a cometary impact. All such impacts, however small the comet mass, involve a huge kinetic energy density (ram pressure) $\rho v_\odot^2/2\approx 10^{15}$ erg/cm$^3$. This is $\sim 10^{10}$ times the thermal energy density (pressure) of the photosphere or the magnetic energy density in a sunspot, being the energy density of a 40 MG magnetic field! Even when the kinetic energy is converted by ablation and ram pressure to heat and kinetic energy of explosion, it is initially spread over only a few scale heights. Hence the explosive airburst energy density of a comet like a Shoemaker-Levy 9 ($10^{15}$ g) is about $10^6$ erg/cm$^3$ equivalent to that of a 5 kG field and still more than the thermal energy density of the photosphere. Thus, while the total energy of most impacting sun-grazers is small compared to that of large flares and CMEs their energy density is so high that local disruption of magnetic fields and triggering of larger scale events are not impossible.
\subsection{Conclusions} Our simple analytic treatment of comet nucleus sublimation gives results for mass loss in reasonable agreement with previous numerical simulations . We have proved that nuclei reaching $p=q/R_\odot \precsim r_*/R_\odot\approx 1.01$ undergo intense bombardment by solar atmospheric ions (energy 2 keV per nucleon) the rate increasing exponentially with depth on a scale of around 500 km. It follows that, above depth $r_*$ where $n=n_*=2.5\times 10^{11}$/cm$^3$, evaporative destruction is essentially by sublimation only. On the other hand, the behavior of any nucleus surviving sublimation to below this height is dominated by atmospheric collisional effects - ablation and ram pressure driven explosion. This creates an exploding air burst followed by a fireball spreading and rising through the atmosphere similarly to the Shoemaker Levy 9 impacts on Jupiter (e.g. Carlson et al. 1997).

Because of the exponential distribution of atmospheric density and the small atmospheric scale height $H$ (compared to $R_\odot$) the terminal heights of these cometary flares is only logarithmically sensitive to the incoming cometary mass (and to the density and latent heat of the nucleus). It only ranges from the upper chromosphere for small to moderate comet masses with typically shallow entry angles to the low chromosphere for very massive steep entry comets.

 The ablated matter (atoms and small particles) decelerates very quickly in collisions with the atmosphere, depositing its kinetic energy and momentum over a few scale heights in tens of seconds, heating the debris and local solar plasma to X-ray emitting temperatures similar to solar flares and with comparable plasma masses and emission measures. Observing such very small impulsive but very luminous 'cometary flares' poses a fascinating challenge but will enable new diagnostics of cometary element abundances from the highly non-solar flash emission spectra. In addition the downward momentum impulse should generate cometary sunquakes as observable ripples in the photosphere akin to those found in conventional magnetically-energized solar flares. In the case of impacts by the most massive comets ($10^{20}$ g or so) the cometary flare energy release ($2\times 10^{35}$ erg) is much larger than that of the largest solar flares ever observed. An impact of this magnitude would have very significant terrestrial effects.

\acknowledgements  We wish to thank referee Paul Weissman for his very detailed reports. These helped us improve the clarity and terminology of the paper and to fill a serious gap in our original discussion of the physics, namely the role of atmospheric ram pressure in destruction of impacting comets, well known to be important for planetary impacts.The paper has also benefited from discussions with K. Battams, R.Galloway, H.S. Hudson, D.W. Hughes,  S.W. McIntosh, C. Schrijver, P. St. Hilaire, and the late B.G. Marsden, to all of whom we are very grateful. We gratefully acknowledge financial support from the UK STFC and RSE Cormack Bequest and from NCAR and TCD.

\end{document}